\documentclass[article]{elsarticle}

\usepackage{graphicx}
\usepackage{float} 
\usepackage{subfigure}
\usepackage{lscape}

\usepackage{listings}
\usepackage{color}

\definecolor{dkgreen}{rgb}{0,0.6,0}
\definecolor{gray}{rgb}{0.5,0.5,0.5}
\definecolor{mauve}{rgb}{0.58,0,0.82}
\definecolor{gray}{rgb}{0.4,0.4,0.4}
\definecolor{darkblue}{rgb}{0.0,0.0,0.6}
\definecolor{lightblue}{rgb}{0.0,0.0,0.9}
\definecolor{cyan}{rgb}{0.0,0.6,0.6}
\definecolor{darkred}{rgb}{0.6,0.0,0.0}

\lstdefinelanguage{XML}
{
  morestring=[s][\color{mauve}]{"}{"},
  morestring=[s][\color{black}]{>}{<},
  morecomment=[s]{<?}{?>},
  morecomment=[s][\color{dkgreen}]{<!--}{-->},
  stringstyle=\color{black},
  identifierstyle=\color{lightblue},
  keywordstyle=\color{red},
  morekeywords={xmlns,xsi,noNamespaceSchemaLocation,type,id,x,y,source,target,version,tool,transRef,roleRef,objective,eventually}
}

\journal{Astronomy and Computing}

\begin{document}

\begin{frontmatter}

\title{IVOA HiPS Implementation in the Framework of WorldWide Telescope}


\author[mainaddress,secondaryaddress]{Yunfei Xu}

\author[mainaddress]{Chenzhou Cui\corref{correspondingauthor}}
\cortext[correspondingauthor]{Corresponding author}
\ead{ccz@nao.cas.cn}
\author[mainaddress]{Dongwei Fan}
\author[mainaddress]{Shanshan Li}
\author[mainaddress]{Changhua Li}
\author[mainaddress]{Jun Han}
\author[mainaddress]{Linying Mi}
\author[mainaddress]{Boliang He}
\author[mainaddress]{Hanxi Yang}
\author[mainaddress]{Yihan Tao}
\author[mainaddress]{Sisi Yang}
\author[mainaddress]{Lan He}
\address[mainaddress]{National Astronomical Observatories, Chinese Academy of Sciences,Beijing, China}
\address[secondaryaddress]{University of Chinese Academy of Sciences, Beijing, China}

\begin{abstract}
The WorldWide Telescope(WWT) is a scientific visualization platform which can browse deep space images, star catalogs, and planetary remote sensing data from different observation facilities in a three-dimensional virtual scene. First launched and then open-sourced by Microsoft Research, the WWT is now managed by the American Astronomical Society (AAS). Hierarchical Progressive Survey (HiPS) is an astronomical data release scheme proposed by Centre de Données astronomiques de Strasbourg (CDS) and has been accepted as a recommendation by International Virtual Observatory Alliance (IVOA). The HiPS solution has been adopted widely by many astronomical institutions for data release. Since WWT selected Hierarchical Triangular Mesh (HTM) as the standard for data visualization in the early stage of development, data released by HiPS cannot be visualized in WWT, which significantly limits the application of WWT. This paper introduces the implementation method for HiPS dataset visualization in WWT, and introduces HiPS data projection, mesh rendering, and data index implementation in WWT. Taking Chang'E-2 lunar probe data as an example, this paper introduces how to convert planetary remote sensing data into a HiPS dataset and integrate it into WWT. This paper also compares the efficiency and memory consumption of WWT loading its native data and HiPS data, and illustrates the application of HiPS in scientific data visualization and science education in WWT.
\end{abstract}


\end{frontmatter}

\section{Introduction}

The WorldWide Telescope (WWT) \cite{worldwidetelescope} is a scientific data visualization platform. It combines astronomical data from dozens of telescopes around the world, allowing users to explore all-sky surveys across the electromagnetic spectrum. WWT provides several different models to visualize astronomical data. In solar system mode, users can view the relative positional relationship between the planets and major satellites in the solar system through three-dimensional scenes. In planetary mode, the surface and elevation high-resolution remote sensing images of each planet and major satellites can be browsed in a way like watching a tellurion. In sky mode, users can view multi-band observation data from different sky survey projects, such as data from the Sloan Digital Sky Survey (SDSS) \cite{blanton2017sloan}, Wide-field Infrared Survey Explorer (WISE) \cite{wright2010wide}, Fermi Space Telescope Gamma-Ray Survey \cite{atwood2009large}, Planck Cosmic Microwave Background Radiation Survey \cite{tauber2004planck}, etc. During browsing, the data can be viewed at any position by dragging and dropping the mouse, and zoom in and out to display at different resolutions, providing users with a seamless, immersive browsing experience. Another of WWT's unique features is its 'Guided Tour'. The tour is a kind of recorded path to which narration, text, and imagery can be added. Users can record or view tours through WWT 3D environments with these 'tours'. Users can share their stories, outreach specialists can create programming, and researchers can distill their discoveries to other researchers or the public \cite{rosenfield2018aas}.

WWT was launched in 2008 by Microsoft Research and then open-sourced in 2015 \cite{wwtopensource}. The American Astronomical Society subsequently took up the project's direction and management. Microsoft Research has developed a series of techniques to achieve a gradual astronomical data browsing method. These techniques use the Hierarchical Triangular Mesh (HTM) \cite{szalay2007indexing} method of SDSS to implement a sphere progressively, and use the Tessellated Octahedral Adaptive Subdivision Transform (TOAST) projection to render plane data on the sphere. However, there are fewer software and libraries supporting HTM other than WWT. SDSS offers HTM development packages in C++ and Java, as well as an extension for Microsoft SQL Server, but all of them have not been updated for a long time (last updated in 2003). The number of HTM-related projects on Github are within ten and have few followers. For PostgreSQL, a widely used relational database management software for astronomical data, HTM does not provide plug-ins or extensions for it, which makes it challenging to use HTM as an indexing method on PostgreSQL for data retrieval. As a result, the HTM method has not been widely used in the astronomical community. Consequently, it is difficult for WWT to integrate the latest released astronomical data. The IVOA (International Virtual Observation Alliance) endorsed the Hierarchical Progressive Survey (HiPS) scheme as an IVOA Recommendation \cite{fernique2017hips}. HiPS is proposed by Centre de Données astronomiques de Strasbourg (CDS), and it has been widely used in the release of astronomical data. HiPS is a hierarchical scheme for the description, storage, and access of astronomical survey data. A "Hierarchical Progressive Survey" allows dedicated clients or browser tools to access and display an astronomical survey progressively, based on the principle that "the more you zoom in on a particular area, the more details show up" \cite{fernique2015hierarchical}. HiPS has evolved into a complete ecosystem. There are more than 20 HiPS nodes in the world, and eight independent clients have been generated. Many astronomical research institutions have chosen to use HiPS to release data, including CDS, ESA, NASA, China-VO and JAXA. More than 550 datasets have been released using HiPS. There are also a dozen of application libraries for developers to use.

By implementing HiPS in WWT, existing HiPS datasets can be browsed in WWT, thereby greatly expanding the WWT data source. Ordinary users can also access the latest astronomical research results through WWT. With the HiPS-supported feature, WWT has further promoted equal access to data by all users, thereby helping to democratize astronomical data and knowledge. This article introduces the method of loading and displaying HiPS data in WWT, based on the HiPS standard version 1.0. Section 2 presents the structure of WWT's 3D rendering framework and describes the main differences between WWT native data and HiPS data. Section 3 introduces details of our method, including how to realize HiPS projection, create HiPS mesh and implement HiPS data organization. In Section 4, we take the Chang'E-2 lunar surface remote sensing data \cite{zhao2011overall} as an example, to show how to convert astronomical observation data into HiPS data and integrate such data into WWT for users to browse. Section 5 is the discussion section. We benchmark the efficiency and memory consumption for displaying HiPS and native data in WWT. Also, we discuss how this work will benefit the WWT application in astronomy education. Section 6 is the summary.

\section{WWT Visualization Framework }
The core of WWT is the 3D rendering framework. Based on this framework, the visualization of astronomical data in a 3D scene is achieved by constructing a mesh model of a sphere, projecting the image data of sky surveys or planets as textures.

\begin{figure}[H]
\centering 
\includegraphics[width=0.7\textwidth]{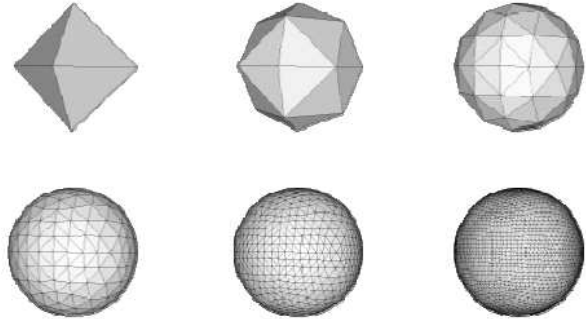} 
\caption{HTM models a sphere recursively}
\label{Fig.1}
\end{figure}

In the visualization process, WWT first draws a spherical model to simulate planets and celestial spheres. To meet the conditions of progressive browsing, this spherical model must be able to implement hierarchical progressive mesh subdivision. WWT uses the HTM to create the spherical model. HTM is a representation of a sphere proposed by astronomers in the Sloan Digital Sky Survey. It first models the sphere as an octahedron, then recursively subdivides each triangular face until the sphere is closely approximated by a highly tessellated polyhedron \cite{szalay2007indexing}, as shown in Fig. \ref{Fig.1}.

\begin{figure}[H] 
\centering
\includegraphics[width=1\textwidth]{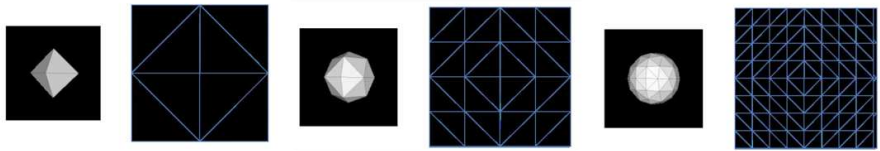}
\caption{The first three levels of HTM subdivision. As each side is subdivided, the new point would be on the surface of a perfect sphere. Even after just three iterations, the polyhedron is getting close to approximating a sphere} 
\label{Fig.2}
\end{figure}

\begin{figure}[H] 
\centering 
\includegraphics[width=0.7\textwidth]{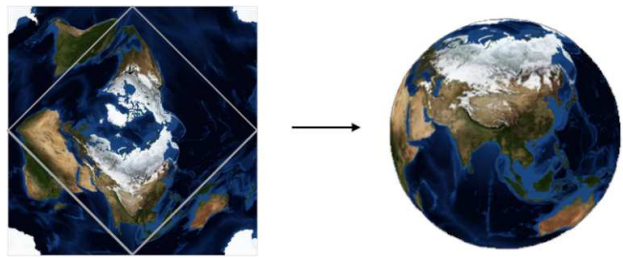}
\caption{TOAST projection-mapped earth}
\label{Fig.3}
\end{figure}

Both sky survey data or planetary remote sensing data take the form of a flat image. To texture images on a spherical model in a 3D scene, a projection conversion is required. WWT uses the TOAST projection which is an extension of the concepts of HTM. TOAST models a sphere as a polygon (starting with an octahedron) that is made up of triangular faces (Fig. \ref{Fig.2}) and folds it flat. Then the north pole would be at the center, the south pole at each of the four corners, and the equator would form a diamond within the square, as shown in Fig. \ref{Fig.3}.

As the data volume of sky survey data and planetary remote sensing data is enormous, WWT organizes data as an image pyramid, in which each lower level contains a higher-resolution version of the total image. One level down has twice the width and twice the height (so four times a greater area) of the image in the previous level \cite{toastprojection}. The level and tiles which should be displayed are judged according to the user's field of view. Then data are attached as textures to the corresponding grid of the spherical model.

\section{Methods}
\subsection{Differences between the visualization of WWT native dataset and HiPS dataset}

As a scheme for the description, storage and access of astronomical data, HiPS has its own data projection, sphere modeling, and data organization standard. These three are completely different from the implementation of the WWT visualization framework. The data projection in WWT is TOAST, while that of HiPS is HEALPix (Hierarchical Equal Area isoLatitude Pixelization of a sphere). The mesh of WWT to create a celestial sphere in a 3D scene is HTM, and HiPS uses a HEALPix rhombic mesh. In terms of data organization, WWT adopts HTM-based data indexing method, while HiPS adopts a HEALPix indexing method. Therefore, to implement loading and rendering of HiPS data in WWT, it is necessary to bridge these differences.

\subsection{Data Projection in the HiPS Scheme}

\begin{figure}[H]
\centering
\includegraphics[width=0.7\textwidth]{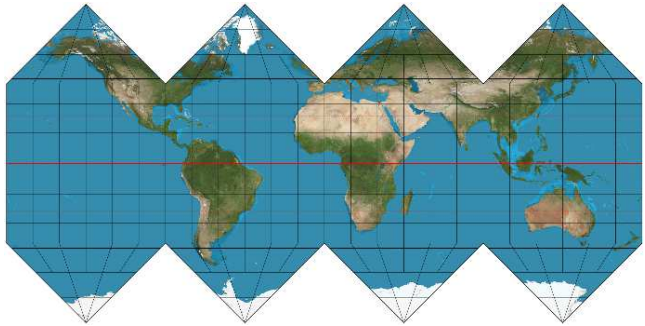}
\caption{HEALPix projection-mapped earth}
\label{Fig.4}
\end{figure}

The HEALPix projection employed in HiPS is a combination of a cylindrical equal-area projection in the equatorial region and an interrupted collignon projection in the polar regions \cite{calabretta2007mapping}. The HEALPix partitioning of the sphere uses a base resolution that divides the sphere into 12 quadrilateral pixels, which are recursively divided into four self-similar smaller pixels in a $2 \times 2$ pattern to create higher order meshes. Projection is achieved by transforming sphere pixels into planes. Fig. \ref{Fig.4} shows the first level of the HEALPix projection, which is described as interrupted, equal area, pseudo-cylindrical projections.

\begin{figure}[H]
\centering
\includegraphics[width=1\textwidth]{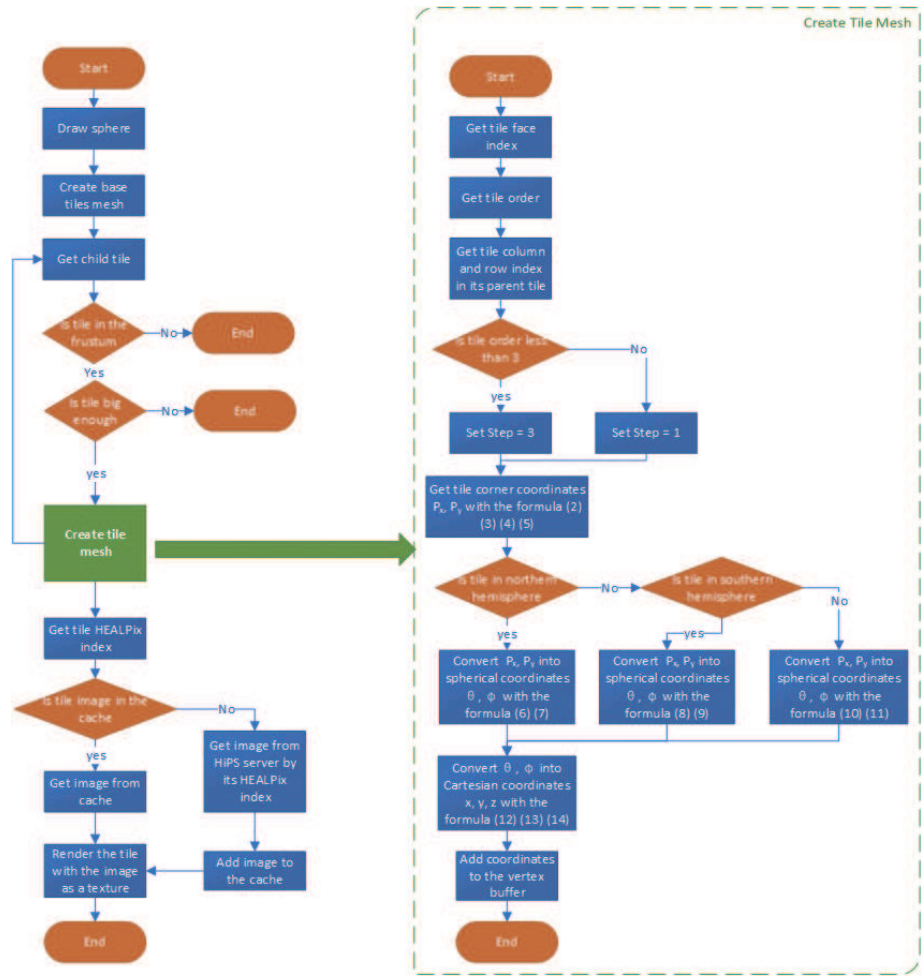}
\caption{The left is the flowchart displaying HiPS data tiles in WWT UI, and the right is the process of creating HEALPix tile meshes.}
\label{Fig.5}
\end{figure}

Due to the substantial differences between TOAST and HEALPix, it is difficult to convert a HEALPix projection into a TOAST projection. We choose to additionally implement the HEALPix projection in the WWT framework to make it coexists with TOAST. In WWT, all data tiles are visualized by displaying them as textures on the celestial sphere mesh. Therefore, the implementation of HEALPix projection is achieved by creating the HEALPix tile meshes. Fig. \ref{Fig.5} is the flowchart which describes the process of displaying HiPS data tiles in WWT UI according to the user's view. The core part is the process of creating HEALPix tile meshes, which is described in a separate flowchart in Fig. \ref{Fig.5}.

\subsection{Creating HEALPix tile meshes in WWT}

\begin{figure}[H]
\centering
\includegraphics[width=0.7\textwidth]{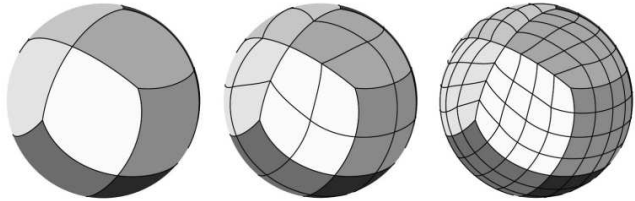}
\caption{HEALPix tile meshes from order 0-2}
\label{Fig.6}
\end{figure}

As described above, HEALPix divides the sphere into 12 quadrilateral zones, and recursively divides each of them into four smaller zones, as shown in Fig. \ref{Fig.6}. The resolution of the HEALPix mesh at a given order $N$ is defined by $N_{side}$, which is the number of divisions of the base resolution tiles and which doubles at each successive order, so that a HEALPix map of $N_{side} = N$ consists of $12N^{2}$ tiles. All tiles at a given order have an equal area of $\Omega_{pix} = \pi/(3N^{2})$, and the centers of the tiles are arranged in rings of equal latitude.

WWT visualization framework is designed solely around triangles and combining triangles to make shapes. The reason is that triangles can be positioned to form just about any shape imaginable. Therefore, to create the HEALPix mesh, we need to define the triangles which can form the mesh, that is, the vertex coordinates of these triangles in 3D space. Instead of defining every corner of every triangle in the mesh, we create a list which contains the coordinates of each vertex, as well as what order they go in.

\begin{figure}[H]
\centering
\includegraphics[width=0.7\textwidth]{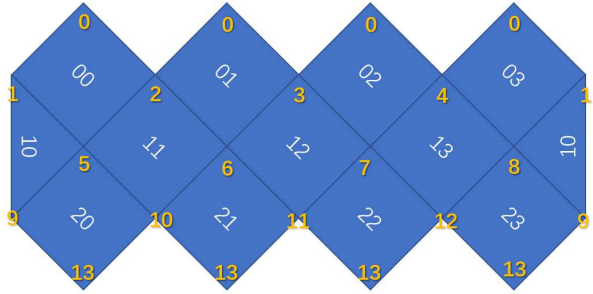}
\caption{The 12 tiles at order 0. The number in the center of each tile is the face index, and the yellow number in the corner is vertex index of each face}
\label{Fig.7}
\end{figure}

The triangle vertices are calculated sequentially for the 12 faces. Two parameters are very important: the $N_{side}$ and the $face index$. The $N_{side}$ is the square root of the number of tiles divided by a face in order $N (N>0)$, e.g. $N_{side} = 2^N$. As its name suggests, the $face index$ is the index of these 12 faces, with a range of $0-11$. The 12 tiles at order 0 are arranged as shown in Fig .\ref{Fig.7}, totaling three rows and four columns, and given a two-digit code according to the order in the row and column as its $face index$. For example, the $face index$ of the second row and the third column is coded as 12, which is 9 in decimal. All child tiles of each face inherit the face number of their parent tile.

When the hierarchy progresses to the next order, each tile of the current order is divided into four tiles, numbered as
\begin{equation}
tile_{index}=parent_{tile_{index}}\times 4+2\times column+row
\end{equation}

To simulate the celestial sphere more finely, extra vertices are needed to be added when the order is lower than 4. To this end, we added a variable \textit{step} to represent the number of vertices added to each side of a tile. When the order is lower than 4, the \textit{step} is set to 3, otherwise set it to 1.

\begin{figure}[H]
\centering
\includegraphics[width=0.7\textwidth]{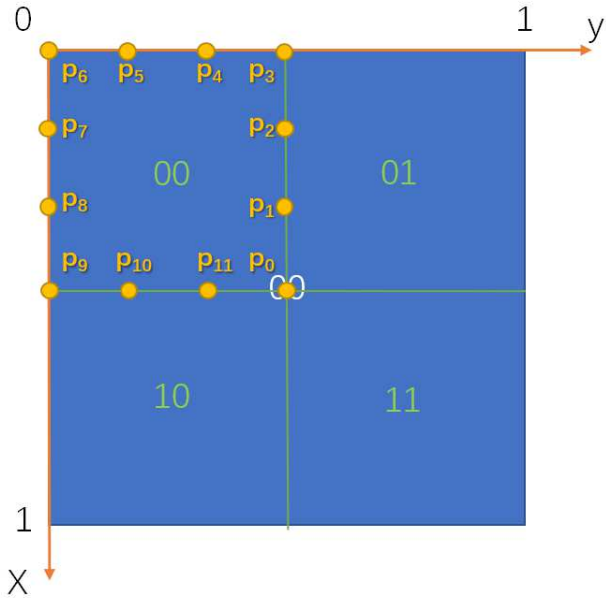}
\caption{Four quads of the face 0, where $p_0 - p_{11}$ are the points of the edges of one quad, and the \textit{step} is 3}
\label{Fig.8}
\end{figure}

In the actual calculation, each tile is divided into 4 quads. Take calculating the coordinates of vertexes in one face at order 0 as an example. As shown in Fig. \ref{Fig.8}, face 0 is divided into four quads. Set the edge length of each face to 1, then the location of the center point coordinate for each quad is $\left ( \frac{1+x}{2N_{side}},\frac{1+y}{2N_{side}} \right )$, where $x$, $y$ is the row index and column index number of the quad respectively. Then the four-corner coordinates are increased or decreased by $dc = \frac{1}{2\times N_{side}}$ on this basis. When the $step$ is not 1, e.g. the quad has extra vertices, another offset $\frac{1}{step\times N_{side}}$ is necessary to be increased or decreased. Then the coordinates of points on a quad can be obtained through the following formula, where $i \in \left [ 0,4step-1 \right ]$.

\begin{equation}
\left ( P_{i_x}, P_{i_y} \right )=\left ( \frac{1+x}{N_{side}}-\frac{i}{step\times N_{side}}, \frac{1+y}{N_{side}} \right )
\end{equation}
\begin{equation}
\left ( P_{(i+step)_x}, P_{(i+step)_y} \right )=\left ( \frac{x}{N_{side}}, \frac{1+y}{N_{side}}-\frac{i}{step\times N_{side}} \right )
\end{equation}
\begin{equation}
\left ( P_{(i+2step)_x}, P_{(i+2step)_y} \right )=\left ( \frac{x}{N_{side}}+\frac{i}{step\times N_{side}}, \frac{1+y}{N_{side}} \right )
\end{equation}
\begin{equation}
\left ( P_{(i+3step)_x}, P_{(i+3step)_y} \right )=\left ( \frac{1+x}{N_{side}}, \frac{y}{N_{side}}+\frac{i}{step\times N_{side}} \right )
\end{equation}

The locations of tile centers on the sphere are defined by $ \left ( z\equiv \cos \theta, \phi  \right )$, where $\theta \in \left [ 0, \pi \right ]$ is the colatitude in radians measured from the north pole and $ \phi \in \left [ 0, 2\pi \right ]$ is the longitude in radians measured eastward. By using these relative coordinates, combined with the face on which they are located, the $z$ and $ \phi$ of each vertex in the spherical coordinate system can be obtained.

\begin{figure}[H]
\centering
\includegraphics[width=0.7\textwidth]{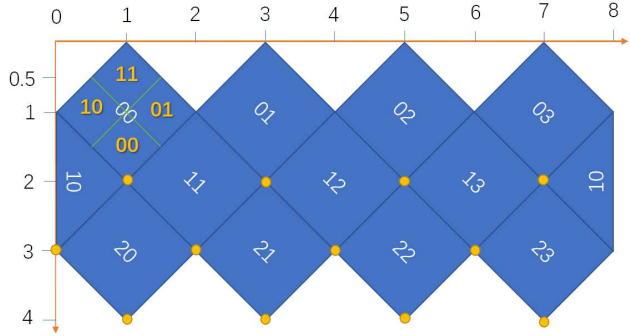}
\caption{Yellow points are the lower vertex of each face in HEALPix order 0}
\label{Fig.9}
\end{figure}

As shown in Fig. \ref{Fig.9}, the coordinates of the lower vertex of each face are set to (2,1), (2,3), (2,5), (2,7), (3,0), (3,2), (3,4), (3,6), (4,1), (4,3), (4,5), (4,7). When the quad is in the 4 faces of the northern hemisphere $\left( face \in \left[0,1,2,3 \right ] \right )$, the calculation formula of a coordinate is:
\begin{equation}
z=1-\frac{{\left( 2-p_x-p_y \right )}^2}{3}
\end{equation}
\begin{equation}
\phi =\frac{\pi}{2}\left (2face+1+\frac{p_x-p_y}{2-p_x-p_y}  \right )
\end{equation}

When the quad is in the four faces of the southern hemisphere $\left(face \in\left [8,9,10,11 \right]\right )$, the formula is:
\begin{equation}
z=\frac{\left ( p_x+p_y \right )^2}{3}-1
\end{equation}
\begin{equation}
\phi = \frac{\pi}{2}\left ( 2face-15+\frac{p_x-p_y}{p_x+p_y} \right )
\end{equation}

When the quad is in the face of the equator $\left( face \in \left [ 4,5,6,7 \right] \right)$, the formula is:
\begin{equation}
z=\frac{2\left ( p_x+p_y-1 \right )}{3}
\end{equation}
\begin{equation}
\phi = \frac{\pi}{2}\left ( 2face-8+p_x-p_y \right )
\end{equation}

Then converting $z, \phi$ to coordinates of the Cartesian coordinate system with the following formula, the coordinates of each vertex in 3D space can be obtained, where $z=\cos \theta$.

\begin{equation}
x=\sqrt{\left ( 1-z \right )\left ( 1+z \right )}\cos\phi
\end{equation}

\begin{equation}
y=\sqrt{\left ( 1-z \right )\left ( 1+z \right )}\sin\phi
\end{equation}
\begin{equation}
z=z
\end{equation}

\begin{figure}[H]
\centering
\includegraphics[width=0.5\textwidth]{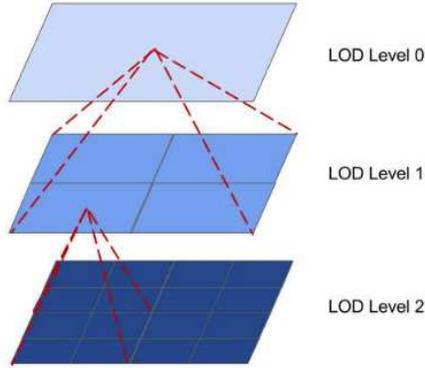}
\caption{ Level of details in one tile of HEALPix mesh from order (level) 0-3}
\label{Fig.10}
\end{figure}

After the coordinates are calculated, the HEALPix mesh can be rendered in the graphics rendering pipeline by adding coordinates to the vertex buffer. In the implementation, the mesh is rendered from order 0, and then the sub-level mesh is successively rendered recursively. The process is the so-called level of details (LOD) implementation. As shown in Fig. \ref{Fig.10}, each tile is subdivided into four tiles in the next order, but it is not always subdivided. The condition of recursion is whether the next order of the current tile is within the camera's field of view, and whether the display size of the tile is large enough in the 3D scene. The mesh of the tile will be constructed only if both conditions are met.

\begin{figure}[H]
\centering
\includegraphics[width=0.7\textwidth]{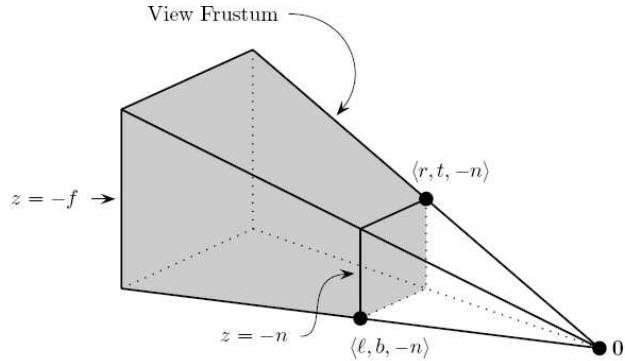}
\caption{ The structure of a view frustum. O is the viewpoint, and ($r, t, -n$) is the upper-right corner's coordinate, ($l, b, -n$) is the lower-left corner's coordinate, such that both of them are located in the near clipping plane.}
\label{Fig.11}
\end{figure}

In a 3D scene, whether an object is displayed on the screen is determined by if it is in the view frustum, which is a cube composed of 6 faces. In order to simulate the real world view, the 3D scene projection is generally on the screen using perspective projection, so the cube is a frustum, as shown in Fig. \ref{Fig.11}. When an object exists in the view frustum, the render pipeline renders it, otherwise it is culled.

The frustum's viewing volume is defined by the parameters: ($left, bottom, -near$) and ($right, top, -near$) which specify the ($x, y, z$) coordinates of the lower-left and upper-right corners of the near clipping plane. Parameters $near$ and $far$ give the distances from the viewpoint to the near and far clipping planes. They should always be positive \cite{viewfrustum}.

Any point $\left (x, y, z \right)$ lying inside the frustum is projected to a point $\left ({x}', {y}', {z}' \right)$ on the near clipping plane $z = -n$ with $l \leq  {x}'\leq  r$ and $b \leq  {y}' \leq  t$. This is the same as
\begin{equation}
1\leq \frac{n\cdot x}{-z}\leq r
\end{equation}
\begin{equation}
b\leq \frac{n\cdot y}{-z}\leq t
\end{equation}

The bounding sphere of a tile is used to determine whether the tile is on the front, back, or intersection of each face of the view frustum. We first obtain the sphere's center and radius, then calculate the distance from the center of the sphere to each face. If the distance is smaller than the radius of the sphere, the tile is intersected by the face. If it is greater and positive, the tile is on the front side of the face. If it is negative, the tile is on the back of the face. The distance calculation formula is as follows, where $C$ is the center coordinate of the sphere, $N$ is the normal vector of the face, and $D$ is the distance from the origin of the coordinate system to the face.

\begin{equation}
distance=\left (C\cdot N  \right )+D
\end{equation}

If a tile is inside the view frustum, it is also necessary to determine whether its size is large enough to fit on the screen. From the corner coordinates of a tile, we can get its screen coordinates, and then get the length of its four sides on the screen. When the longest side is less than a predetermined value $L$, the tile is considered too small to display on the screen in the current field of view. The value of $L$ is empirical. If this value is small, more tiles will be displayed on the screen, resulting in a more exquisite display and more memory consumption. On the contrary, the display accuracy is reduced, and the memory is used less.

\subsection{Implementing the HiPS Data Organization}

After completing the rendering of HEALPix tile meshes, the HiPS data are used as a texture mapped on a specific tile. In the HiPS data organization, the image data are segmented according to the grid and hierarchy of HEALPix. For example, the image of the all-sky survey data is cut into 768 tiles at order 3, one for each HEALPix grid, i.e., the image segmentation of the all-sky survey can be mapped to the grid of HEALPix mesh one-to-one by the HEALPix index number. HEALPix provides two numbering schemes for tiles index, namely the RING scheme and NESTED scheme. In RING scheme tiles are counted down from north to south along each iso-latitude ring. The NESTED scheme arranges the tiles into 12 tree structures concerning their base-resolution tiles. HiPS chooses the NESTED scheme as its index scheme \cite{anoverviewofhips}.

\begin{figure}[H]
\centering
\includegraphics[width=0.7\textwidth]{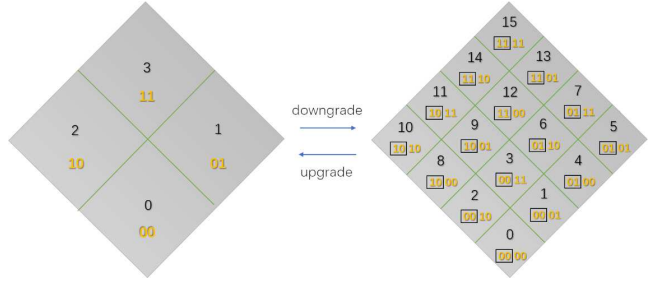}
\caption{Downgrade and upgrade of the tile index with the Nested schema}
\label{Fig.12}
\end{figure}

Consider the Nested index $p_n \in \left [ 0,12N_{side}^2-1 \right ]$, and define ${p_n}'= \left( p_n mod N_{side}^2 \right )$, where ${p_n}'$ denotes the nested tile index within each base-resolution element. The binary representation of ${p_n}'$ is $...b_3b_2b_1b_0$. Given the grid resolution parameter $N_{side}$, the location of a tile on each base resolution tile is represented by the two indices $x$ and $y \in \left\{0, N_{side}-1 \right\}$, and their starting point is in the base-resolution tile, with the $x$ index running along the North-East direction, while the $y$ index runs along the North-West direction. The binary representation of ${p_n}'$ determines the values of $x$, and $y$ as the following combinations of even and odd bits, respectively, $x=...b_2b_0$, $y=...b_3b_1$ \cite{gorski2005healpix}. When the hierarchy is subdivided, the new ${p_n}'$ can be obtained by splicing the $XY$ index of the parent order and that of the new order, as shown in Fig. \ref{Fig.12}.

A HiPS dataset is generally published via a classical HTTP server. A HiPS browser is able to directly access to the tiles from the server, without requiring any database access for localizing the data \cite{anoverviewofhips}. By the method we described above, the WWT has the ability to get HEALPix tile indices covering the user's view, then what we need to do is access the corresponding tiles by their URLs from the HiPS service.

All tiles of a HiPS dataset are stored in a collection of directories. The structure of these directories follows the simple hierarchy: $order \rightarrow  tiles$, by using respectively the prefix $Norder$ for orders, and $Npix$ for tiles. To avoid creating too large directories, the tiles are grouped by 10, 000 items, using the subdirectory prefix $Dir$. Therefore, the URL of a specific HiPS tile is shaped like $http://\{domain\}/\{data\_name\}/Norder\{0\}/Dir\{1\}/Npix\{2\}$. All the contents in braces are variables, where $\left\{domain \right\}$ is the domain name that provides the HiPS service, and $\left\{data\_name\right\}$ is the name of the specific HiPS dataset, $Norder\left\{0\right\}$ indicates the order of the tile, $Dir\left\{1\right\}$ is the folder where the tile is located, and $Npix\left\{2\right\}$ is the index of the tile at its order.

In the process of generating the HEALPix tile meshes, the HEALPix indices of tiles to be displayed have been calculated from the user's view. Images corresponding to the indices are requested and added to the download queue. After images are downloaded from the HiPS server, they are rendered as textures on corresponding tiles to complete the data display.

\section{Application}
\subsection{Deployment HiPS dataset in WWT}

We extend the data configuration file of WWT by adding the HiPS information to it, then WWT has the ability to load the HiPS dataset automatically. The configuration file is an XML file, the format of which is as follows:
\lstset{language=XML}
\begin{lstlisting}
<ImageSet Generic="False" DataSetType="HiPS_SkyMap"
 BandPass="Visible" Name="DSS2 Blue (XJ+S)"
 Url="http://alasky.u-strasbg.fr/DSS/DSS2-blue-XJ-S/Norder{0}/Dir{1}/Npix{2}"
 BaseTileLevel="0" TileLevels="9"
 FileType=".jpg"
 Projection="Healpix"
>
</ImageSet>
\end{lstlisting}

The primary information is in the $ImageSet$ element, where the $DataSetType$ attribute indicates the data type, and all HiPS dataset have a $HiPS$ prefix to distinguish them from WWT's native data. The $Url$ attribute is the access address of the dataset, and its format is as described in the previous section. The $BaseTileLevel$ attribute indicates that the data loading starts from the first order. Due to the WWT loading the data recursively, all data are loaded from order 0. Despite CDS having updated all their HiPS services by adding the low orders (Norder 0-2) data in early 2019, and encouraging other HiPS creators to do the same update, there are still some HiPS services that have not yet provided low orders of data. For HiPS services without low orders data, a blank and transparent image is loaded as the texture of a tile at low orders. The $TileLevels$ attribute represents the total number of orders of the dataset. The $FileType$ attribute represents the file format. The $Projection$ attribute indicates that the dataset is in the HEALPix projection.

\subsection{Integrating the Chang'E-2 lunar surface data in WWT}

China's Chang'E-2 Lunar probe acquired 7 meter-resolution digital orthophoto map (DOM) data of the whole moon surface. These data have now been published \cite{change2datarelease}. By processing the data and publishing the data as a HiPS dataset, it can be browsed through the WWT planetary mode.

The publicly released DOM data are divided into two parts: the polar region data and the equatorial region data. The former cover the range from 70 degrees north latitude to the north pole and 70 degrees south latitude to the south pole, and the data projection is Polar Azimuth projection. The equatorial region data cover between 70 degrees north latitude and 70 degrees south latitude, in Mercator projection. For the convenience of downloading and reading, all of the data are divided into 844 pieces, and each piece of data volume is around 1 GB. The data are stored in tiff format, and each piece of data is accompanied by a .prj file and a .tfw file for recording the projection and its geographic coordinates.

Creating a HiPS dataset with these data requires conversion of the data format and projection. Since we use the tools released by CDS to make HiPS dataset, the first step is to convert the data format and merge the .tiff, .prj, and .tfw files into a single FITS (Flexible Image Transport System) file, then create the HiPS dataset using all the FITS files.

In the conversion of the data to the FITS format, the corresponding FITS header keywords are generated based on the information in the .prj and .tfw files. The keyword 'WCS' represents 25 kinds of projections, including the Mercator projection and Polar Azimuth projection used in these data, which are represented by 'MER' and 'STG' in the FITS header, respectively. However, the HiPS generation tool provided by CDS does not support Mercator projection, so it is necessary to convert all non-polar regions data into Polar Azimuth projection, which is the same as the projection of data in polar regions. The conversion is implemented using the open-source geographic data processing library GDAL \cite{gdal}.

The HiPSGen creates HiPS data from FITS files by three processes: index calculation, image segmentation, and image conversion. In the index calculation process, each piece of an image is divided into several blocks according to the size of $1024\times1024$. Then HEALPix indices at the max order of each block are calculated according to its corner coordinates, where the max order of the Chang'E-2 dataset is ten based on the image resolution. The coordinates of the center point and four corners of each block are also calculated in the process. All the information is written into an intermediate file in the "HPXFinder" folder for the next image segmentation process.

\begin{figure}[H]
\centering
\includegraphics[width=0.5\textwidth]{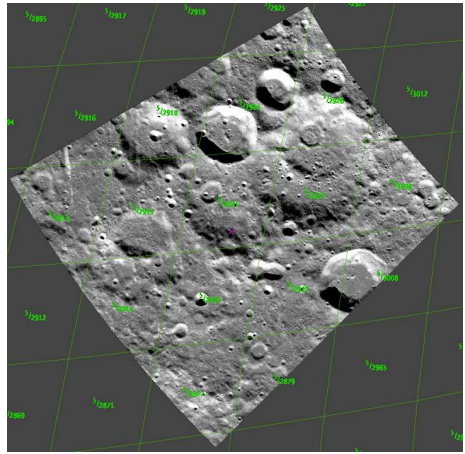}
\caption{HEALPix Grid under Polar Azimuth projection}
\label{Fig.13}
\end{figure}

The projection conversion process is achieved by rearranging the pixels of each image according to the HEALPix index. Take a tile at order 10 as an example. Its size is $512\times512$ and each pixel of it is equivalent to a HEALPix tile at order 19 $(10+(\log_{2}512))$. The HEALPix index at order 19 is obtained by the $RA$ and $Dec$ of each pixel of the origin image, then the $XY$ coordinates at the new image are calculated according to the HEALPix index. Finally, we fill in the pixel value of each point in the original image to the corresponding location of the new image. If the value is empty, it is set to null, as shown in Fig.\ref{Fig.13}.

In this way, the conversion of the original image from Polar Azimuth projection to HEALPix projection is completed along with the image segmentation. CDS has developed a batch conversion tool $HiPSGen$ \cite{hipsgen} to do this work. The HiPS properties file also can be automatically generated by the tool.

\begin{figure}[H]
\centering
\includegraphics[width=1\textwidth]{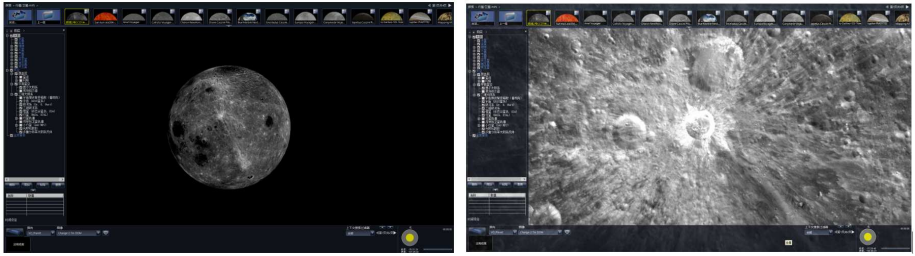}
\caption{Visualizing the Chang'E-2 DOM HiPS dataset in WWT}
\label{Fig.14}
\end{figure}

The produced HiPS data are published by a web server and can be accessed through each tile's URL. After adding the HiPS information in WWT's configuration file, we can browse the data in WWT, as shown in Fig.\ref{Fig.14}.

\section{Discussion}
\subsection{Benchmarking the display efficiency and memory consumption}

HiPS datasets in WWT can also be used to make guided tours. They operate in the same way as to make tours using WWT native data (TOAST data). There is no additional learning cost for tour creators. On tour playback, the tour using HiPS dataset takes less time to load, but its memory consumption will be more massive. We select 12 datasets, 6 of which are HiPS data and the other 6 are TOAST data, and compare the time cost and memory consumption of rendering in WWT between HiPS data and TOAST data. The results are shown in Table \ref{tab:table1}, where $Max Order$ refers to the highest order of the data. It should be noted that due to different splitting methods, the maximum order of HiPS data and TOAST data is not consistent, and the resolution of most TOAST data is lower than that of HiPS dataset. The most commonly used resolution for HiPS tiles is $512\times512$, while the resolution of the TOAST tile is $256\times256$. $Ti$ represents the time it takes for tiles at order 0 to fill the entire display screen. In the test environment, the resolution of the display is $2560\times1440$. $Mi$ is the system memory consumed by the WWT when tiles at order 0 fill the entire display screen. $Tm$ refers to the time it takes for the max order tiles to fill the entire screen. $Mm$ refers to the system memory consumed when the max order data fills the entire screen. We can see that in our method the HiPS has a disadvantage in memory consumption. The leading cause is that the most used tile size of HiPS data is twice the size of TOAST data. We also compared the performance using HiPS data in $256\times256$ size, which is identical as the size of TOAST data. The result shows the loading time and memory consumption are significantly improved when the tile size is reduced. There is an advantage in HiPS data in terms of maximum order data loading speed. The reason is that HiPS takes less tiles to fill the entire screen as it has a higher resolution per tile. These test results may vary if different HiPS or TOAST tile resolutions were applied, and data providers may need to consider the relative importance of memory consumption and tile loading time when deciding which tile resolution to use.

\subsection{Impact on the application of astronomy education based on WWT}

\begin{figure}[H]
\centering
\includegraphics[width=1\textwidth]{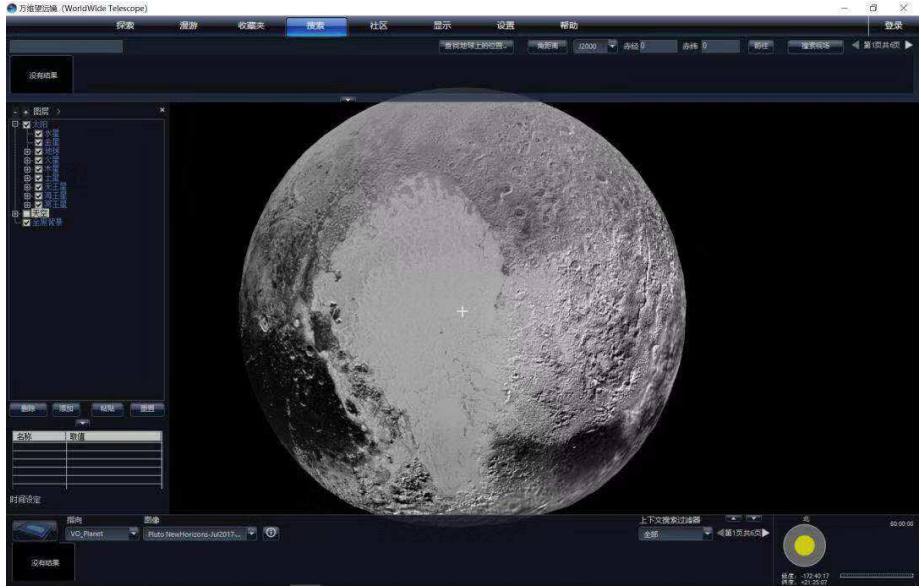}
\caption{"Heart of Pluto" from New Horizons in WWT }
\label{Fig.15}
\end{figure}

The HiPS supporting feature is not only very critical to WWT itself but also important to the education and public outreach activities based on this platform. As mentioned before, this special platform can provide tons of real astronomical data from observatories around the world in a convenient way, and allows people to interact with data using the "Guided Tour" feature \cite{rosenfield2018aas}. With the orbit data visualization function, 3D model insertion, geographic information data visualization and many other unique features, WWT has more potential in education and public outreach than other similar software. In China, with the ten years of development and promotion by Chinese Virtual Observatory (China-VO), nearly 40 schools and science museum had already used WWT to teach science class or public outreach. Although gaining a lot of recognitions from astronomical community in China in recent years, problems gradually emerged. Very little data have been updated due to the development difficulties and the data loading speed, which lags far behind the average web browsing speed. Users were not able to learn the latest astronomical data they saw in the news and have to deal with the relatively low speed. As an application platform, these shortcomings were fatal because users may lose interest immediately when they cannot find what they want. With the new HiPS technology add in, these problems were solved to some extent. The latest data, including data from famous astronomical projects like Gaia \cite{brown2018gaia}, Chang'E, New Horizons \cite{stern2009new}, etc. These projects usually have great influence on the public and amateur astronomers. For example, the "Heart of Pluto" from New Horizons aroused public interest and repercussions worldwide because of its special morphological and characteristics. It became a good science education opportunity and did attract a certain degree of social attention. Unfortunately, the New Horizons data were missing on the WWT platform for a long time, until now. With the method in this paper, all the datasets implemented into the platform can be used not only for browsing but also for "Guided Tour" design, as shown in Fig. \ref{Fig.15}. Meanwhile, the loading speed was improved and this was definitely good news for educators who want to use WWT in the classroom.

The AAS has released a new version of the WorldWide Telescope Desktop (based on Windows OS) at the AAS 235 meeting \cite{aasworldwidetelescope}. This version integrated our work and introduced the HiPS-supporting feature. Meanwhile, the web client of WWT will also be updated to support HiPS as well for tours and data to interoperate. However, many users cannot upgrade or are reluctant to upgrade WWT to the latest version, which makes tours made with HiPS datasets unable to play on these users' clients. This is a problem we need to solve in the next stage. We hope the WWT community can expand its reach to let users understand the benefits of HiPS datasets, giving them the incentive to upgrade the WWT version.

\section{Summary}

This paper presented an implementation of integrating IVOA HiPS into the WWT framework. WWT is astronomical visualization software that can directly link to all levels of scientific and education. HiPS is increasingly being used for the release of astronomical data, and more than 550 datasets are released using HiPS so far. Implementing HiPS data visualization in WWT requires the realization of HEALPix projection, mesh creating, and data organization. This article details the implementation of these key technologies.

We also made the 7 meter-resolution DOM data acquired by China's Chang'E 2 Lunar probe as a HiPS dataset, and introduced the method for integrating it into WWT, which can be used as a reference for integrating future planetary remote sensing data into WWT.

We compared the efficiency and memory consumption of WWT loading its native data and HiPS data. HiPS has an advantage in display efficiency, but the memory consumption is more massive, which is caused by the larger file size of each tile of HiPS dataset.

Based on the work of this paper, China-VO has already announced an official release of the new version of WWT with HiPS technology in May 2019, 39 selected HiPS Sky maps and 49 selected HiPS planet maps have been added into China-VO version WWT. In August 2019, a WWT teacher training program was held in Hangzhou. More than 60 teachers from different regions of China participated in this training and experienced many of new astronomical data sets with joy. The practice has proved that the implementation of HiPS technology is also of great help to astronomical science education and public outreach and to the forward-looking concept data-driven astronomy education and public outreach (DAEPO) \cite{cui2018iau}.

\section*{Acknowledgements}
This work is supported by National Natural Science Foundation of China (NSFC)(11803055), the Joint Research Fund in Astronomy (U1731125, U1731243, U1931132) under cooperative agreement between the NSFC and Chinese Academy of Sciences (CAS). We acknowledge the Centre de Données astronomiques de Strasbourg (CDS) for expert advice and help for the use of HiPS, in particular, Pierre Fernique, Thomas Boch, and Mark Allen. We would also like to express appreciation to Jonathan Fay of Microsoft Research for his valuable comments. Data resources are supported by National Astronomical Data Center (NADC) and Chinese Virtual Observatory (China-VO). This work is supported by Astronomical Big Data Joint Research Center, co-founded by National Astronomical Observatories, Chinese Academy of Sciences, and Alibaba Cloud.


\bibliography{mybibfile}

\begin{landscape}
\begin{table}[]
\caption{The comparison of data loading speed and memory consumption between the loading of WWT native data and HiPS data in WWT. In which Ti refers to the time of loading
the initial order, Tm refers to the time of loading the max order, Mi refers to the memory consumption of the initial order, Mm refers to the memory consumption of the max order. Please note that all HTM data tile is $256\times256$ in size, and HiPS data is $512\times512$ except for PLANCK.}
\label{tab:table1}
\begin{tabular}{llllllll}
\hline
\textbf{\begin{tabular}[c]{@{}l@{}}Data\\ name\end{tabular}}                                                     & \textbf{\begin{tabular}[c]{@{}l@{}}Data\\ type\end{tabular}} & \textbf{\begin{tabular}[c]{@{}l@{}}Ti\end{tabular}} & \textbf{\begin{tabular}[c]{@{}l@{}}Tm\end{tabular}} & \textbf{\begin{tabular}[c]{@{}l@{}}Max\\ order\end{tabular}} & \textbf{\begin{tabular}[c]{@{}l@{}}Mi\end{tabular}} & \textbf{\begin{tabular}[c]{@{}l@{}}Mm\end{tabular}} & \textbf{\begin{tabular}[c]{@{}l@{}}Tile\\ Size\end{tabular}} \\ \hline
Digitized Sky Survey                                                                                             & \begin{tabular}[c]{@{}l@{}}TOAST-\\ Skymap\end{tabular}        & 2.28s                                                                                     & 7.73s                                                                                  & 12                                                           & 851920KB                         & 825584KB    &   $256\times256$                                                                                                                                                   \\ \cline{2-8}
                                                                                                                 & \begin{tabular}[c]{@{}l@{}}HiPS-\\ Skymap\end{tabular}       & 2.25s                                                                                     & 8.60s                                                                                  & 9                                                            & 1327612KB                        & 1488840KB   &   $512\times512$                                                                                                                                                  \\ \hline
\begin{tabular}[c]{@{}l@{}}WISE\\ All Sky\end{tabular}                                                           & \begin{tabular}[c]{@{}l@{}}TOAST-\\ Skymap\end{tabular}        & 1.90s                                                                                     & 9.58s                                                                                  & 7                                                            & 821484KB                         & 821832KB    &  $256\times256$                                                                                                                                                   \\ \cline{2-8}
                                                                                                                 & \begin{tabular}[c]{@{}l@{}}HiPS-\\ Skymap\end{tabular}       & 2.10s                                                                                     & 3.86s                                                                                  & 8                                                            & 1551080KB                        & 1221216KB   &   $512\times512$                                                                                                                                                     \\ \hline
\begin{tabular}[c]{@{}l@{}}2Mass:\\ Imagery\end{tabular}                                                         & \begin{tabular}[c]{@{}l@{}}TOAST-\\ Skymap\end{tabular}        & 2.23s                                                                                     & 11.86s                                                                                 & 8                                                            & 864448KB                         & 821052KB    &  $256\times256$                                                                                                                                                   \\ \cline{2-8}
                                                                                                                 & \begin{tabular}[c]{@{}l@{}}HiPS-\\ Skymap\end{tabular}       & 2.30s                                                                                     & 4.83s                                                                                  & 9                                                            & 1335512KB                        & 1306756KB   &   $512\times512$                                                                                                                                                     \\ \hline
\begin{tabular}[c]{@{}l@{}}Moon:\\  \\ Lunar Reconnaissance\\ Orbiter WAC \\ Global Morphologic Map\end{tabular} & \begin{tabular}[c]{@{}l@{}}TOAST-\\ Planet\end{tabular}        & 2.12s                                                                                     & 16.66s                                                                                 & 10                                                           & 817988KB                         & 840520KB    &  $256\times256$                                                                                                                                                   \\ \cline{2-8}
                                                                                                                 & \begin{tabular}[c]{@{}l@{}}HiPS-\\ Planet\end{tabular}       & 1.65s                                                                                     & 1.53s                                                                                  & 6                                                            & 1413816KB                        & 23987602KB  &   $512\times512$                                                                                                                                                   \\ \hline
\begin{tabular}[c]{@{}l@{}}Jupiter:\\  \\ PIA07782, Cassini's \\ Best Maps of Jupiter\end{tabular}               & \begin{tabular}[c]{@{}l@{}}TOAST-\\ Planet\end{tabular}        & 1.43s                                                                                     & 4.46s                                                                                  & 4                                                            & 846060KB                         & 831084KB    &  $256\times256$                                                                                                                                                                                                                                                                                                        \\ \cline{2-8}
                                                                                                                 & \begin{tabular}[c]{@{}l@{}}HiPS-\\ Planet\end{tabular}       & 2.30s                                                                                     & 0.98s                                                                                  & 3                                                            & 1756340KB                        & 1083836KB   &   $512\times512$                                                                                                                                                     \\ \hline
\begin{tabular}[c]{@{}l@{}}PLANCK \end{tabular}               & \begin{tabular}[c]{@{}l@{}}TOAST-\\ Skymap\end{tabular}        & 0.71s                                                                                     & 0.54s                                                                                  & 6                                                            & 990892KB                         & 1000896KB    &  $256\times256$                                                                                                                                                                                                                                                                                                       \\ \cline{2-8}
                                                                                                                 & \begin{tabular}[c]{@{}l@{}}HiPS-\\ Skymap\end{tabular}       & 0.63s                                                                                     & 0.59s                                                                                  & 3                                                            & 757152KB                        & 712348KB   &  $256\times256$                                                                                                                                                                                                                                                                                                       \\ \hline
\end{tabular}
\end{table}
\end{landscape}

\end{document}